\documentclass{custom_frontiers}

\usepackage{graphicx}
\usepackage[latin1]{inputenc}
\usepackage{amsmath}
\usepackage[sort&compress]{natbib}
\usepackage{rotating}
\usepackage{fixltx2e}
\usepackage{siunitx}
\usepackage{amssymb}
\usepackage{gensymb}

\begin{document}

\title{Switched-Capacitor Realization of Presynaptic Short-Term-Plasticity and Stop-Learning Synapses in 28~nm CMOS}
\author{Marko Noack$^{1,*}$, Johannes Partzsch$^1$, Christian Mayr$^2$, Stefan H\"anzsche$^1$, Stefan Scholze$^1$, Sebastian H\"oppner$^1$, Georg Ellguth$^1$,and Rene Sch\"uffny$^1$\\
$^1$ Chair of Highly Parallel VLSI Systems and Neuromorphic Circuits, Technische Universit\"at Dresden, Dresden, Germany\\
$^2$ Institute of Neuroinformatics, University of Zurich and ETH Zurich, Zurich, Switzerland\\
$^*$ Corresponding author: Marko Noack, marko.noack@tu-dresden.de}
\maketitle

\begin{abstract}
Synaptic dynamics, such as long- and short-term plasticity, play an important role in the complexity and biological realism achievable when running neural networks on a neuromorphic IC. For example, they endow the IC with an ability to adapt and learn from its environment.
In order to achieve the millisecond to second time constants required for these synaptic dynamics, analog subthreshold circuits are usually employed. However, due to process variation and leakage problems, it is almost impossible to port these types of circuits to modern sub-100nm technologies. 
In contrast, we present a neuromorphic system in a 28~nm CMOS process that employs switched capacitor (SC) circuits to implement 128 short term plasticity presynapses as well as 8192 stop-learning synapses. The neuromorphic system consumes an area of 0.36~mm$^2$ and runs at a power consumption of 1.9~mW.
The circuit makes use of a technique for minimizing leakage effects allowing for real-time operation with time constants up to
several seconds.
Since we rely on SC techniques for all calculations, the system is composed of only generic mixed-signal building blocks. These generic building blocks make the system easy to port between technologies and the large digital circuit part inherent in an SC system benefits fully from technology scaling. 
\end{abstract}

\markright{Synaptic Dynamics in 28 nm CMOS}
\pagestyle{myheadings}

\section{Introduction}

Biological synapses employ a range of plasticity mechanisms in modulating their stimulus transmission.
For example short-term plasticity on the timescale of hundreds of milliseconds has been identified as a crucial constituent of dynamic neural information processing, allowing for temporal filtering \citep{grande05}, selective information transmission \citep{mayr09a} and pattern classification in attractor networks \citep{mejias09}.
Long-term plasticity, with induction on the minute to hour scale, is used for pattern learning \citep{brader07} and topology formation, allowing a network to be structured for solving a particular problem \citep{rubinov11}.
Both of these mechanisms employ exponential time windows with time constants on the order of 10-1000~ms. 

Most analog neuromorphic implementations of plasticity rely on subthreshold circuits \citep{indiveri06} to achieve the small currents necessary for these long time constants. However, these are hard to port to advanced CMOS techologies, since leakage currents rapidly increase with down-scaling, reaching the range of the desired signal currents \citep{roy03}. 
Some plasticity circuits have also been implemented in OTA-C architectures \citep{noack11,koickal07}, but these suffer from the same problems with small currents. Digital plasticity circuits \citep{cassidy11} are not subject to this limitation, but have limited biological veracity due to their digital state variables.
For subthreshold circuits, an additional problem is the increase of device mismatch and process variation \citep{Kinget2005}, making transistors almost unusable for the exponential computation that subthreshold circuits rely upon. This is why even recent subthreshold neuromorphic systems have been manufactured in quite large technologies \citep{bartolozzi07,indiveri10,moradi13}, with the sole exception a recent design in 90~nm \citep{park14}.

The SC technique offers a viable alternative, as it utilizes robust charge-based signal transmission. That is, it computes with charges that are equivalent to accumulating the continuous signal currents of subthreshold circuits across time, thereby raising signal levels compared to the subthreshold approach. This approach has already been successfully applied to neuromorphic neuron implementations \citep{vogelstein07,folowosele09b}.

In this paper we present SC circuits that implement presynaptic adaptation and synaptic plasticity in a 28~nm CMOS process. The short-term (presynaptic) plasticity has been adapted for SC \citep{noack12} from the biology-derived neurotransmitter release model of \cite{markram98}. The long-term (synaptic) plasticity circuit implements the stop learning stochastic synapse model of \cite{brader07}. To the best of our knowledge, this represents the first time the well-known stop-learning paradigm has been translated to SC circuits. Please note: While this paper focusses on dynamics, a companion paper \citep{mayr14c} presents the static neuromorphic components (weight implementation, neurons, etc) and the overall system integration. 

\cite{vogelstein07} and \cite{folowosele09b} have chosen a straightforward SC approach with conventional CMOS switches, as leakage currents were not a concern in their chosen technology nodes. However, this approach is not possible in deep-submicron technologies such as the employed 28~nm process. The leakage for open switches would preclude storing a signal on the required 10-1000~ms timescale. Thus, we describe circuit techniques to reduce leakage currents, in turn allowing us to achieve high time constants. 
The entire neuromorphic system consists of standard analog building blocks and synthesizable digital logic, making it easy to port between technologies. As detailed later, the system architecture has been optimized for mismatch reduction.

\section{Material \& Methods}
\subsection{Overall System}
\label{sec_design_system}

\begin{figure}
\centering
\includegraphics[width=0.48\textwidth]{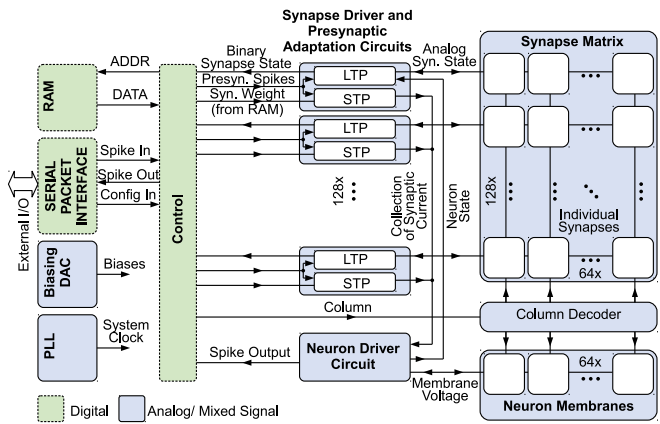}
\caption{\label{fig_overview}Overview of the neuromorphic system with mixed signal SC blocks (e.g. presynaptic adaptation, synapse matrix and neurons), digital control, synaptic weight RAM, biasing DAC, PLL clock input and serial packet IO.}
\end{figure}

\begin{figure}
\centering
\includegraphics[width=0.48\textwidth]{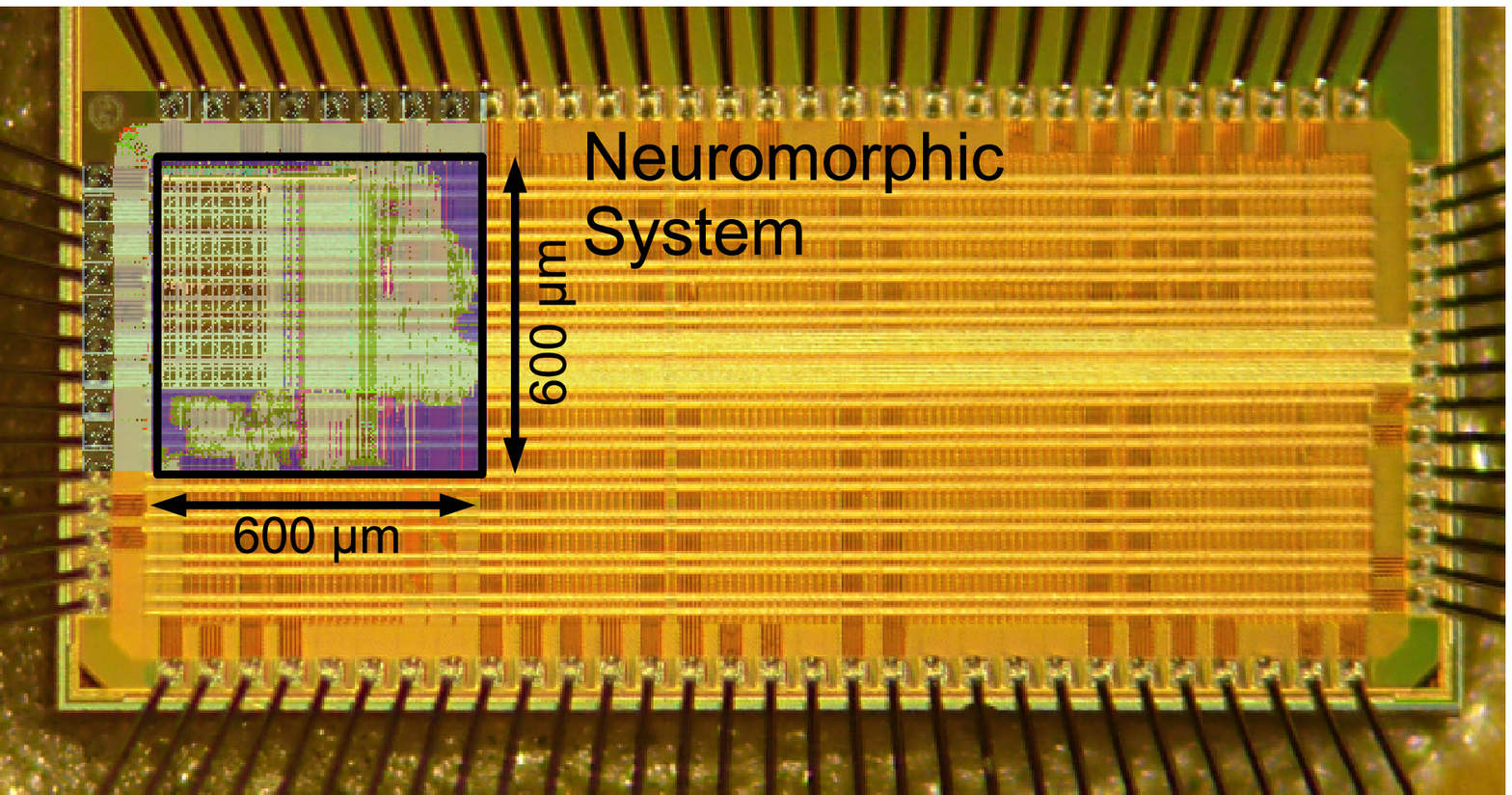}
\caption{\label{fig_die}Chip photograph with overlay of the $\SI{600}{\micro\meter} \times \SI{600}{\micro\meter}$ neuromorphic system layout. Die size is $\SI{1.5}{\milli\meter} \times \SI{3}{\milli\meter}$.}
\end{figure}

Fig. \ref{fig_overview} gives an overview of the system. 128 input circuits at the left side realize presynaptic short-term dynamics for their respective row in the synaptic matrix \citep{noack12}, while the 64 neurons at the bottom are driven by their respective column, providing the output (i.e. stimulation) signal as a function of the 8192 synapses in the system, which couple presynaptic input to neurons. Synaptic weights are stored in a dedicated RAM block separate from the synapse matrix. 

The entire driving circuitry of presynapses, synapses and neurons is situated at the left hand side of the matrix. A state machine cycles through the columns of the synaptic matrix. At the start of the cycle, the input pulses that were registered during the last cycle are forwarded to the driver circuits and the corresponding presynaptic adaptation state is computed. Then, each synaptic column is activated sequentially, and the synaptic plasticity change of a synapse at a specific row is computed based on presynaptic pulse activity of that row and the membrane state of the neuron of the current column. Concurrently, the presynaptic pulses are integrated on the neuron. 
Sharing the active driver circuitry for all neurons respectively for all synapses of a row inherently reduces mismatch effects, as the only remaining mismatch between synapses is the mismatch of their state-holding capacitors. Mismatch between transistors, i.e. between active circuits, is only felt between rows.

The circuit design utilizes only digital core devices of the 28nm SLP (super low power) technology. In contrast to the current biasing usually employed in neuromorphic ICs \citep{yang12}, the neuromorphic SC circuits require voltages provided by a digital-to-analog converter (DAC) to set amplitude parameters such as scaling of presynaptic adaptation, etc. This saves pins and offers an easy and robust configurability.

Time constants are set via counters that govern the switching cycles of the SC circuits. Thus, scaling of the clock frequency effectively scales the speed of the system, keeping the resolution relative to the chosen time base. As the clock speed scaling retains the relative speed of all processes, the same configuration for all parameters (amplitudes and time constants) can be used irrespective of the speed-up, nominally giving the same results.
The neuromorphic system was designed for speeds from biological real-time (corresponding to a 0.62~ms full cycle time of the synaptic matrix) up to an acceleration of 100.

Communication with the system is provided by a JTAG interface, implementing a generic packet-based protocol. Similar to the communication setup in \citep{hartmann10,scholze11a}, these packets contain configuration and incoming/outgoing pulse communication data. Additionally, two configurable test outputs allow for monitoring analog voltages, such as membrane potentials. With its minimal interface, using only 6 signal pins and two bias pins (one bias current and one pin for common mode voltage), the neuromorphic system can be easily integrated into a multi-core system mediated by an FPGA. A chip photograph is shown in Fig. \ref{fig_die}. The neuromorphic system occupies \SI{0.36}{\square\milli\meter} and is surrounded by various test structures. The overall IC has a size of $\SI{1.5}{\milli\meter} \times \SI{3}{\milli\meter}$. 

\subsection{Implementation of Presynaptic Short-Term Plasticity}
\label{sec_presynapse}
\subsubsection{Model}

The presynaptic adaptation circuit implements the model of synaptic dynamics proposed in \cite{noack12}, which is derived from a  model based on biological measurements \citep{markram98}. The major drawback of the original approach in \cite{markram98} with respect to a switched-capacitor implementation is the need for a wide-range voltage multiplier for calculating the product of the facilitation and depression state variables. Existing multipliers are rather complex, very area consuming \citep{Hong1984} or need large operational amplifiers driving resistive loads \citep{Khachab1991}. In contrast, the model proposed in \cite{noack12} is capable of approximately reproducing the original model without any multiplier circuit and with a minimum effort on analog circuitry in general. 

The iterative description of the proposed model is shown in eqs. (\ref{eq:circuit_u}) -- (\ref{eq:circuit_psc}):
\begin{align}
u_{n+1} &= u_n\cdot(1-U)\cdot\mathrm{e}^{-\frac{\Delta t_n}{\tau_{u}}}+U\label{eq:circuit_u}\\
R_{n+1} &= \left((1-\alpha)\cdot R_{n} + \alpha\cdot u_n\right)\cdot\mathrm{e}^{-\frac{\Delta t_n}{\tau_R}}\label{eq:circuit_r}\\
PSC_{n} &= A \cdot ( u_n-R_n)\label{eq:circuit_psc}\,.
\end{align}
It provides the amplitude $PSC_{n}$ of the postsynaptic current for successive presynaptic spikes incorporating their spiking history, where $n$ is the number of the observed spike and $\Delta t_n$ denotes the time between $n$-th and $(n+1)$-th spike. The model is capable of reproducing facilitation and depression as well as various combinations of both mechanisms. Facilitation is modeled by variable $u$, which is adopted from \cite{markram98}. At each incoming presynaptic spike $u$ is increased by a certain amount, depending on $U$. Between spikes it exponentially decays back to $U$ with time constant $\tau_u$. Thus, $u$ is bound to the interval $[U,1]$. Variable $R$ describes the depression mechanism and is also increased at every presynaptic spike. Inspired from \cite{markram98} the amount depends on the current value of $u$. The strength of depression is controlled via $\alpha$, which can be any value between 0 and 1. Between spikes $R$ decays back to 0 with time constant $\tau_R$. The resulting PSC amplitude is then calculated by the difference of $u_n$ and $R_n$, scaled by a factor $A$. The PSC decays with time constant $\tau_{PSC}$.

\subsubsection{Circuit Implementation}
\begin{figure*}
\centering
\includegraphics[width=0.9\textwidth]{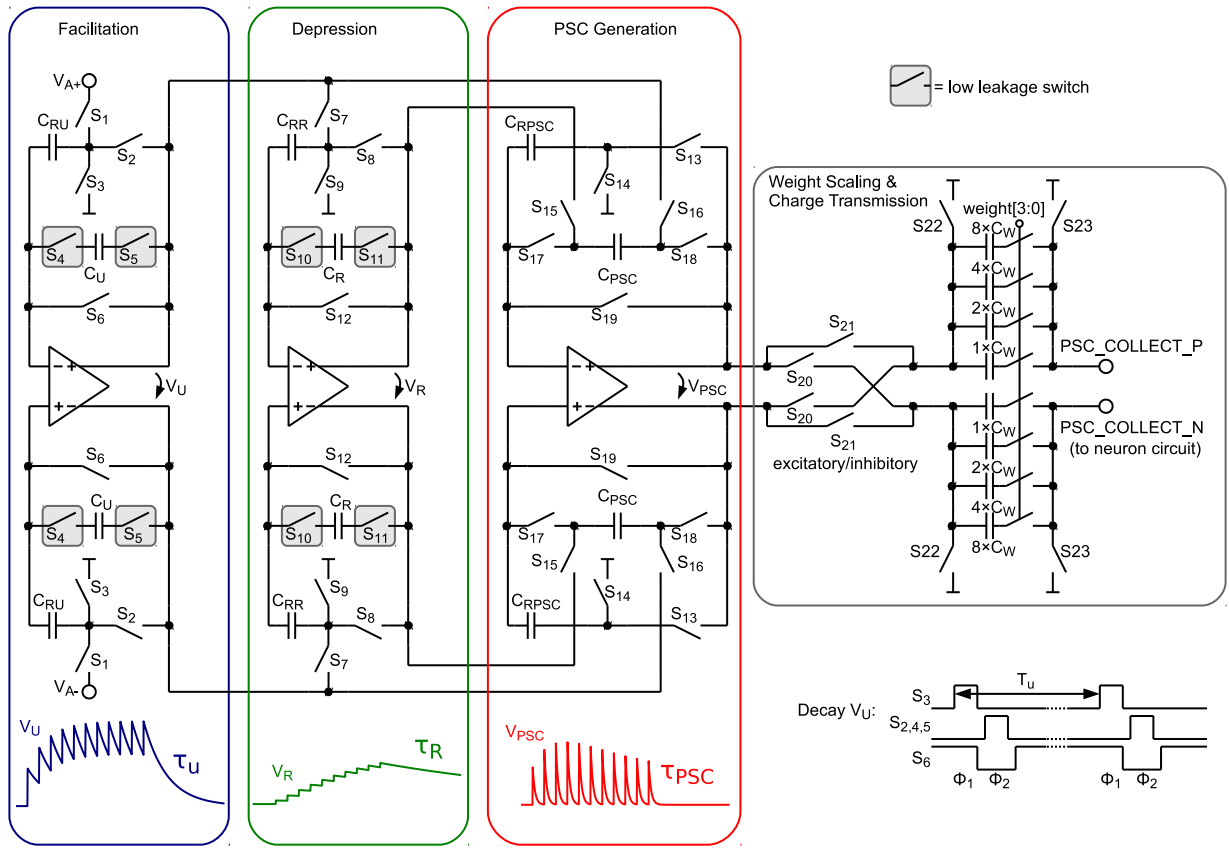}
\caption{\label{fig_circuit}Schematic of the presynaptic adaptation circuit comprising 3 fully-differential SC leaky integrator circuits. Capacitors storing the value of the corresponding model variables are encapsulated by dedicated low-leakage switches.}
\end{figure*}

In order to transform the iterative model to continuous-time, the exponential time dependence can be implemented with exponentially decaying voltage traces. These are generated by the circuit shown in Fig. \ref{fig_circuit} 
for the internal state variables $u$, $R$ and $PSC$, which model facilitation, depression and postsynaptic current trace, respectively. At incoming presynaptic spikes these decay traces are triggered  and the resulting PSC amplitude is calculated by the difference of facilitation and depression value as shown in Eq. \ref{eq:circuit_psc}. In Fig. \ref{fig_circuit} the circuit schematic is shown comprising three similar parts, for calculating $V_U$, $V_R$ and $V_{PSC}$. 

\begin{figure}
\centering
\includegraphics[width=0.48\textwidth]{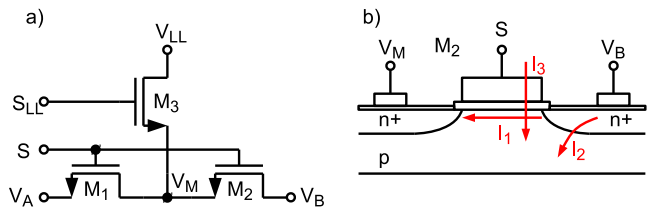}
\caption{\label{fig_lowleakage_switch}(a) Low-leakage switch configuration. (b) Cross-section of MOS Transistor M2 with denoted subthreshold leakage ($I_1$), junction leakage ($I_2$) and gate leakage ($I_3$).}
\end{figure}

\begin{figure}
\centering
\includegraphics[width=0.48\textwidth]{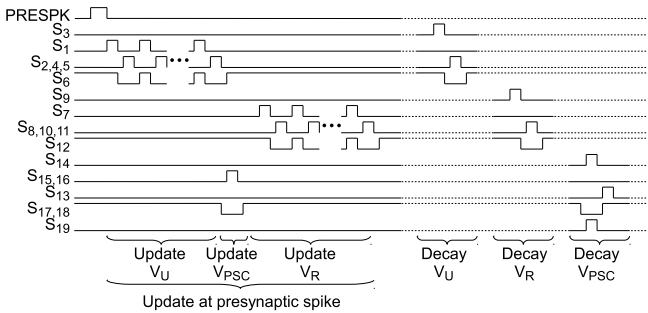}
\caption{\label{fig_switching}Switch signals for update at an incoming presynaptic spike and for exponential decays of $V_U$, $V_R$ and $V_{PSC}$. Dotted lines indicate that decay events can occur independently as well as simultaneously.}
\end{figure}

When a presynaptic spike occurs these voltages are updated by a special switching scheme presented in Fig. \ref{fig_switching}.
$V_U$ is increased towards $V_A$, which represents the global scaling factor $A$ in Eq. \ref{eq:circuit_psc}. The number of switching events of the $V_U$ update determines the parameter $U$. $\alpha$ is set by the number of switching events of the $V_R$ update. Switches $S_{17}$ and $S_{18}$ transfer the voltage difference of $V_U$ and $V_R$ to $V_{PSC}$. 

Between incoming spikes an exponential decay of $V_U$, $V_R$ and $V_{PSC}$ is performed by SC leaky integrator circuits. The working principle will be explained for the facilitation subcircuit and can be applied analogously for depression and PSC generation. On every decay event (see "Decay $V_u$" in Fig. \ref{fig_switching}) $C_{RU}$ (\SI{5}{\femto\farad}) is discharged in a first switching phase $\Phi_1$ (see also bottom right of Fig. \ref{fig_circuit}). In this period $C_U$ (\SI{75}{\femto\farad}), which stores the value of the facilitation variable, is fully decoupled from the circuit. Switching phase $\Phi_2$ performs a charge equalization on $C_U$ and $C_{RU}$. Thus, on every decay event $V_U$ is decreased by a factor $\frac{C_U}{C_U+C_{RU}} = \frac{15}{16}$. These decay events are repeated with period $T_u$. With $\frac{15}{16} = \mathrm{exp}(-\frac{T_u}{\tau_u})$ we can easily calculate $T_u$ for a desired decay time constant $\tau_u$:
\begin{equation}
T_u = -\tau_u \cdot\mathrm{ln}(\frac{15}{16}) \approx \tau_u \cdot 0.0645\,.
\end{equation}
Since $T_u$ is derived from a digital counter driven by the system clock, $\tau_u$ is proportional to the counter size and system clock frequency and allows to set time constants ranging from a few milliseconds to about one second. 
In order to scale the system's overall speed there is a tunable system clock divider, which enables to operate the circuit from biological real-time up to a 100-fold acceleration, keeping all relative timings without the need for adjusting bias voltages. 

With the period of the matrix column cycle, the resulting exponentially decaying PSC voltage is sampled on the 4-bit binary-weighted capacitor $C_W$ and transferred to the neuron circuit.

\subsubsection{Leakage Reduction}
\label{sec_leakage_reduction}
The maximum achievable time constant is limited by subthreshold leakage and junction leakage in the switches (see $I_1$ and $I_2$, resp. in Fig. \ref{fig_lowleakage_switch}b) \citep{roy03}. A dedicated technique similar to \cite{ellguth06} and \cite{ishida06} has been applied for switches surrounding capacitors $C_{U}$ and $C_{R}$ where the switch transistor is split into two transistors (see Fig. \ref{fig_lowleakage_switch}a). If the switch is in off-state the middle node $V_M$ is clamped to a fixed voltage $V_{LL}$. Switch signals $S$ and $S_{LL}$ are non-overlapping. With $V_{LL} = \SI{250}{\milli\volt}$, which is equal to the common-mode voltage, drain-source voltage of M1 and M2 is kept low, which minimizes subthreshold leakage. Furthermore the amount of leakage current is independent of the voltage at the other switch terminal. Junction leakage is minimized by minimal sized drain and source terminals. With a reduced voltage swing of about $V_{DD}/2$ all switches can be implemented with NMOS transistors only, which keeps leakage currents low and reduces circuit complexity. Especially the concept of isolating capacitors by low-leakage switches makes it possible to reach time constants up to \SI{600}{\milli\second}, which is the maximum controllable setting in our design, despite using small capacitance values in the \SI{28}{\nano\meter} technology node (which naturally has high leakage). This is demonstrated by the measurements in Sec. \ref{sec_results_time_constants}. Thus, we achieve an off-resistance of about $\SI{600}{\milli\second}/\SI{75}{\femto\farad}=\SI{8}{\tera\ohm}$, which corresponds to a conductance of \SI{125}{\femto\siemens}. In contrast to another technique recently proposed by \cite{Rovere2014}, which requires two auxiliary low offset opamps, our solution is much more area and power efficient and satisfies our leakage constraints.

\subsubsection{Proposed Opamp}
\label{sec_opamp}
\begin{figure}
\centering
\includegraphics[width=0.48\textwidth]{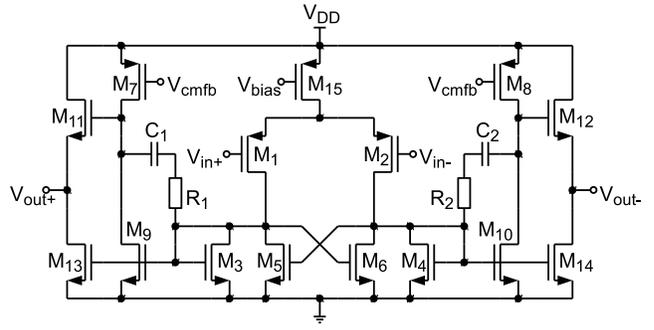}
\caption{\label{fig_opamp_circuit}Proposed opamp circuit used for buffering $V_u$, $V_R$ and $V_{PSC}$.}
\end{figure}

For buffering $V_u$, $V_R$ and $V_{PSC}$ a two stage opamp is used (see Fig. \ref{fig_opamp_circuit}), since transistor stacking is difficult at supply voltages of \SI{1}{\volt}. A gain boosting technique similar to \cite{dessouky00} has been applied, where the load of the first stage has been split into two cross-coupled transistors ($M_3$, $M_5$ and $M_4$, $M_6$). By connecting the gates of $M_5$ and $M_6$ to the opposite output of the first stage a positive feedback is generated.
The common-mode voltage of the first stage is well defined by the diode connected transistors $M_3$ and $M_4$ whereas the common-mode voltage of the output stage ($M_7$ -- $M_{14}$) is controlled by an SC CMFB network. In order to derive stability a classical miller compensation ($C_1$, $R_1$, $C_2$, $R_2$) has been applied using poly resistors and custom designed metal-oxide-metal capacitors. At the output an NMOS source follower ($M_{11}$ -- $M_{14}$) is connected, which enhances slew rate performance. Thus, the output voltage range is limited to \SIrange[range-phrase = -- ]{0}{500}{\milli\volt}, which corresponds to the allowed voltage range of the low-leakage switches. The input common mode voltage range is \SIrange[range-phrase = -- ]{0}{420}{\milli\volt}, which is sufficient for $V_{cm} = \SI{250}{\milli\volt}$. The opamp consumes an area of \SI{68}{\square\micro\meter} and achieves an open-loop gain of \SI{54}{\decibel}. It is designed to operate in biological real-time, as well as in a 100-fold accelerated environment. In fast mode the opamp draws \SI{30}{\micro\watt} of power and has a slew rate of \SI[per-mode=symbol]{60}{\volt\per\micro\second}. As the capacitor settling time scales with speed-up, the power consumption in real-time operation can be reduced by a factor of 100, i.e. down to \SI{300}{\nano\watt}.

\subsubsection{Offset Compensation}
\label{sec_offset_compensation}
Due to the small area occupied by the opamp, which is important for large scale integration, mismatch results in a maximum input offset voltage of about \SI{\pm16}{\milli\volt}. Nevertheless, this offset can be compensated by a simple auto-zeroing technique \citep{enz96}. As can be seen in  Fig. \ref{fig_circuit}, in the sampling phase ($\Phi_1$) input voltages and common-mode voltages, respectively, are sampled against virtual ground of the opamp (switches $S_6$, $S_{12}$ and $S_{19}$ are closed). Since the offset voltage is present at the opamp input at this time, it is also sampled, and thus, canceled out at the output in the second phase ($\Phi_2$). Despite the existence of more advanced auto-zeroing techniques in the literature, this technique has been chosen, because neither additional capacitors nor additional switching phases are required, reducing area and circuit complexity.

\subsection{Switched-Capacitor Implementation of a Bistable Stochastic Synapse}
\label{sec_design_synapse}

\subsubsection{Model}
\label{sec_fusi_model}
The stop learning model of long-term plasticity has been introduced in \cite{brader07}, based on earlier work in \cite{fusi00}. The model represents a synapse with two stable states, potentiated and depressed, whereby the state transition between both stable states is regulated via a continuous internal state X(t) of the synapse. X(t) is influenced by a combination of pre- and postsynaptic activity, namely the presynaptic spike time $t_{pre}$ and the value of the neuron membrane voltage $V_{mem}(t)$.
A presynaptic spike arriving at $t_{pre}$ reads the instantaneous values $V_{mem}(t_{pre})$ and $C(t_{pre})$. The conditions for a change in X depend on these instantaneous values in the following way:

\begin{align}
\mathrm{X} &\rightarrow\mathrm{X}+a &if& & \{V_{mem}(t_{pre})&>\theta_V\phantom{sp} and \label{eq_fusi_a}\\
& & & & \theta_{up}^l <&C(t_{pre})<\theta_{up}^h\}& \notag\\
\mathrm{X} &\rightarrow\mathrm{X}-b & if& & \{V_{mem}(t_{pre})&\leq\theta_V \phantom{sp} and\label{eq_fusi_b}\\
& & & & \theta_{down}^l <&C(t_{pre})<\theta_{down}^h\}\,,& \notag
\end{align}
where a and b are jump sizes and $\theta_V$ is a voltage threshold. In other words, X(t) is increased if $V_{mem}(t)$ is elevated (above $\theta_V$) when the presynaptic spike arrives and decreased when $V_{mem}(t)$ is low at time $t_{pre}$.  The $\theta_{up}^l$, $\theta_{up}^h$, $\theta_{down}^l$ and $\theta_{down}^h$ are thresholds on the calcium variable. The calcium variable C(t) is an auxiliary variable (see \cite{brader07} for details) that provides a low-pass filter of the postsynaptic spikes. This gives the ability to stop the learning based on thresholded, long-term averages of postsynaptic activity. 
In the absence of a presynaptic spike or if stop learning is active (i.e. C(t) hits the respective threshold), then X(t) drifts toward one of two stable values:

\begin{align}
\frac{d\mathrm{X}}{dt} &=\alpha & if& & \mathrm{X}&>\theta_X \label{eq_fusi_alpha}\\
\frac{d\mathrm{X}}{dt} &=-\beta & if& & \mathrm{X}&\leq\theta_X \label{eq_fusi_beta}
\end{align}

The bistable state of the synapse is determined according to whether X(t) lies above or below the threshold $\theta_X$.
Computationally, this model is interesting because through X(t) it can learn a graded response to an input pattern even though the output weight of the synapses is binary. The model also has some biological veracity, being sensitive to pre-post and post-pre spike patterns in a manner similar to the well-known spike time dependent plasticity \citep{brader07}.

\subsubsection{Circuit Implementation}
\label{sec_fusi_circuit}

\begin{figure*}
\centering
\includegraphics[width=0.9\textwidth]{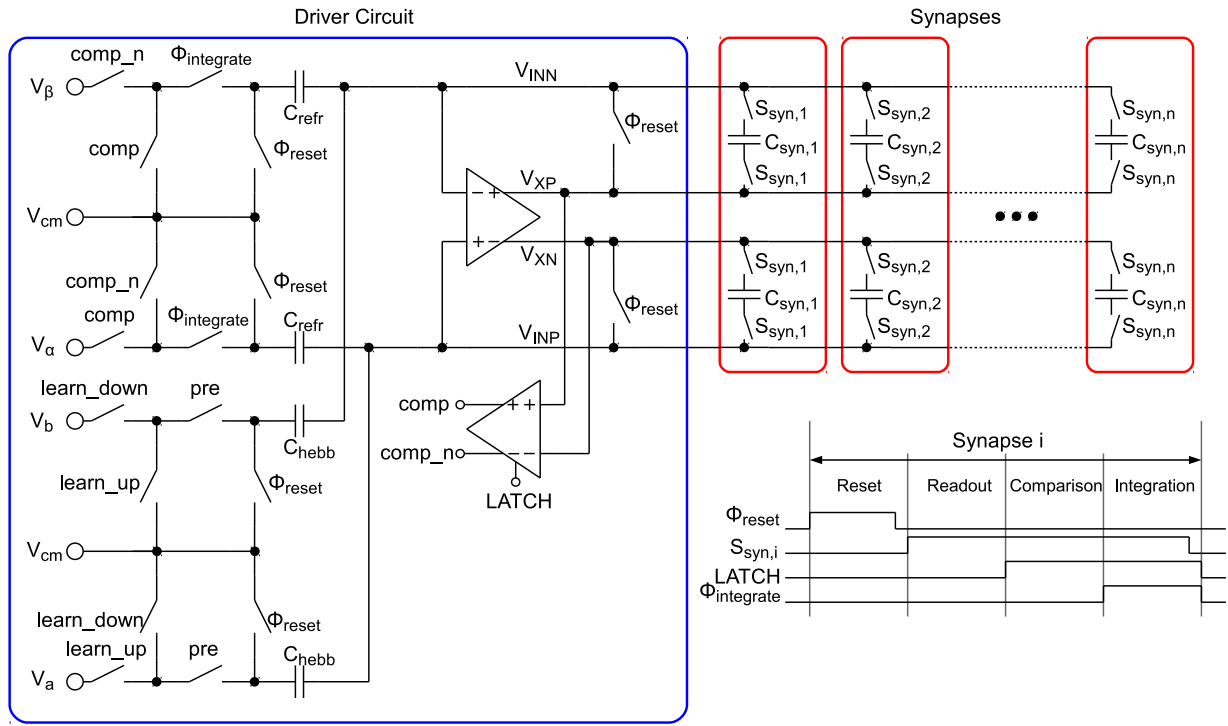}
\caption{\label{fig_ltpcircuit}LTP circuit}
\end{figure*}

The circuit schematic shown in Fig. \ref{fig_ltpcircuit} replicates the model described in Eqs. (\ref{eq_fusi_a}) -- (\ref{eq_fusi_beta}). In contrast to the circuit presented in \cite{indiveri06} our implementation makes use of SC technique. Thus, the model equations are solved in a time-discrete fashion, which enables the use of low-leakage switches as shown in Sec. \ref{sec_leakage_reduction} to achieve very low drift rates $\alpha$ and $\beta$. The time-discretization also allows for time multiplexing the single synapse circuits, thus, one driver circuit (see blue box in Fig. \ref{fig_ltpcircuit}) can drive multiple (in our case 64) synapses (red boxes). Due to the removal of active elements, one synapse circuit can be reduced to only 2 capacitors and 4 low-leakage switches storing the synapse state $X$ (cp. Eqs. (\ref{eq_fusi_a}) -- (\ref{eq_fusi_beta})) as a differential voltage. The synapse occupies an area of $\SI{3.6}{\micro\meter} \times \SI{3.6}{\micro\meter}$ which is shared equally by the two synapse capacitors with \SI{22}{\femto\farad} each. These are custom-designed metal-oxide-metal capacitors, utilizing an interdigitated fingered layout in the complete 5-layer metal stack with cut-outs on the lower two layers for wiring. The low-leakage switches are located directly below the capacitors. Each synapse can be connected to the driver circuit via switches $S_{syn,i}$, where $i$ indicates the column number in the synapse matrix, and 4 wires $V_{INP}$,$V_{INN}$,$V_{XP}$ and $V_{XN}$. The driver circuit is basically an SC integrator, which integrates different voltages $V_\alpha$, $V_\beta$, $V_a$ and $V_b$ in dependence of synapse state, neuron state and incoming presynaptic spikes onto the synapse capacitors $C_{syn,i}$. The integrator's opamp is the same as for the presynaptic driver presented in Sec. \ref{sec_opamp}. As shown in the timing diagram in the lower right corner of Fig. \ref{fig_ltpcircuit}, the operation principle can be divided into 4 phases "Reset", "Readout", "Comparison" and "Integration" for one synapse. All synapses of one row are cycled through sequentially, whereas all rows are processed in parallel.

In the reset phase an offset compensation of the opamp (cp. Sec. \ref{sec_offset_compensation}) is performed, which avoids the integration of a possible offset voltage as well as residual charge on the relatively long wires to the synapses. Therefore switches annotated with $\Phi_{reset}$ are closed, which closes a negative unity-gain feedback loop around the opamp. The offset voltage appearing at the opamp input is then stored on capacitors $C_{refr}$ and $C_{hebb}$ and can be subtracted in the integration phase.

After reset a readout of the synapse state is performed. Switches $S_{syn,i}$ of the currently active synapse $i$ are closed, which places the synapse capacitors in the feedback path of the opamp. The voltage stored on the capacitors, i.e. the synapse state $X$, is now visible at the opamp output between the differential lines $V_{XP}$ and $V_{XN}$.

\begin{figure}
\centering
\includegraphics[width=0.4\textwidth]{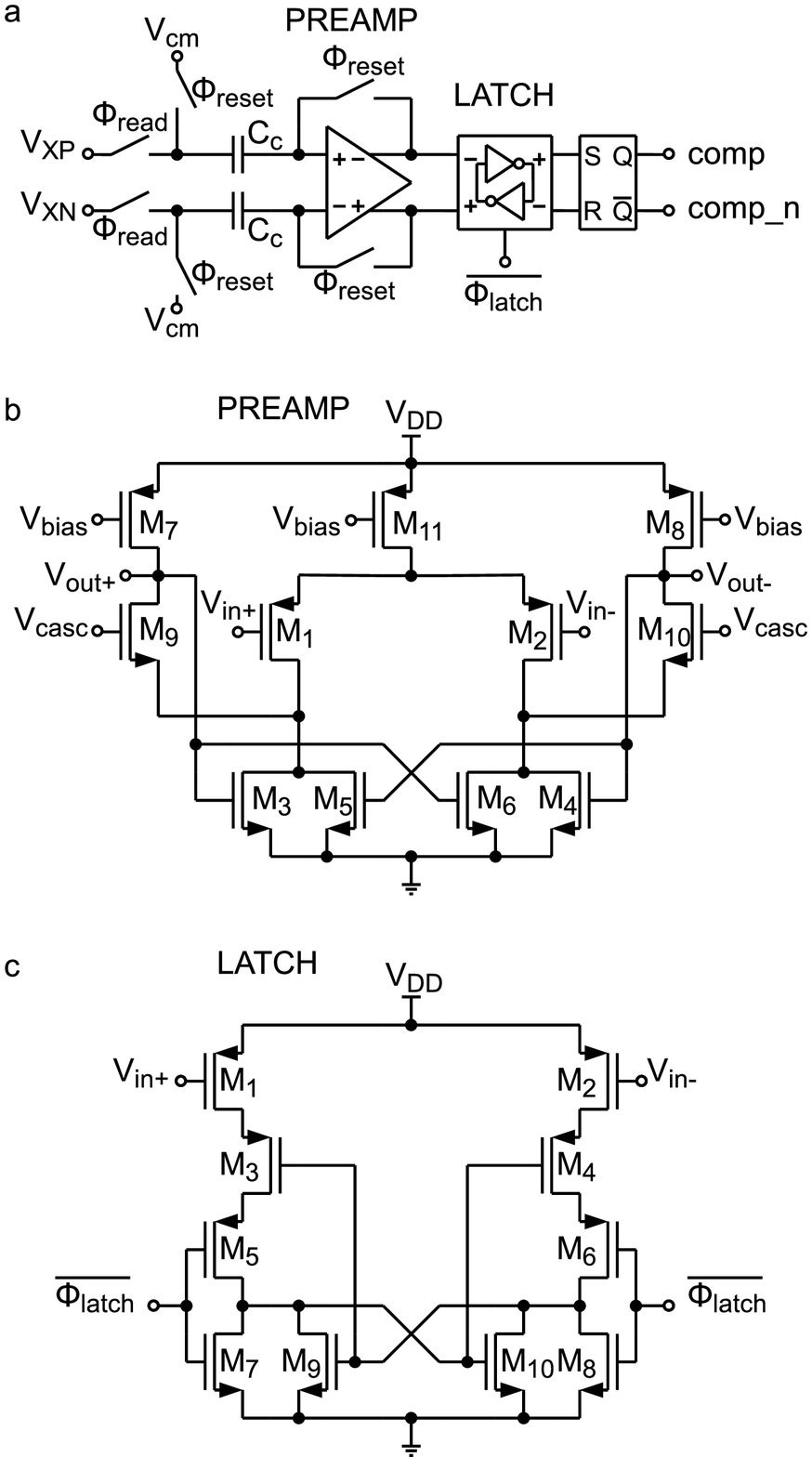}
\caption{\label{fig_compcircuit}a) Comparator circuit with offset-compensated preamplifier, compensation capacitors $C_c$ and latch circuitry. b) Preamplifier circuit schematic. c) Latch circuit schematic.}
\end{figure}

When the readout is completed the synapse capacitors stay connected and a comparison of the synapse state with threshold $\Theta_X$ is performed. In the implementation $\Theta_X$ is fixed at 0.5, thus, the comparator (see Sec. \ref{sec_comparator}) only has to compare whether $V_{XP} > V_{XN}$. After comparison the result is provided by signals comp and its inverted counterpart comp\_n.

In the integration phase the refresh part (see Eqs. (\ref{eq_fusi_alpha}) and (\ref{eq_fusi_beta})) and the hebbian part (Eqs. (\ref{eq_fusi_a}) and (\ref{eq_fusi_b})) of the learning model are performed. In this phase switches annotated with $\Phi_{integrate}$ are closed. If comp is high then the differential synapse voltage $V_X$ is increased by $\frac{C_{refr}}{C_{syn}}\cdot(V_\alpha - V_{cm})$, otherwise it is decreased by $\frac{C_{refr}}{C_{syn}}\cdot(V_\beta - V_{cm})$. This results in refresh rates of 
\begin{equation}
\alpha = \frac{C_{refr}}{C_{syn}} \cdot \frac{(V_\alpha - V_{cm})}{\Delta t}
\end{equation}
and
\begin{equation}
\beta = \frac{C_{refr}}{C_{syn}} \cdot \frac{(V_\beta - V_{cm})}{\Delta t}\,,
\end{equation}
where $\Delta t = \SI{0.62}{\milli\second}$, which is the time needed for processing the 64 synapses of a row sequentially (in biological real-time mode).

If a presynaptic input spike arrives, then switch signal pre is high during the integration phase. In dependence of the postsynaptic membrane state $\Theta_V$ signals learn\_up and learn\_down are set. The neuron circuit providing the membrane state is an SC leaky integrate-and-fire neuron presented in the companion paper \cite{mayr14c}. It is equipped with two comparator circuits for spiking threshold detection and for judging the current membrane state, i.e. the $V_{mem}(t_{pre})\gtrless\theta_V$ condition of Eqn. \ref{eq_fusi_a} resp. Eqn. \ref{eq_fusi_b}. If $V_{mem}(t_{pre})>\theta_V$, then learn\_up is high and learn\_down is low (neglecting the "stop learning" mechanism for now). Thus, the upward jump size is calculated by
\begin{equation}
a = \frac{C_{hebb}}{C_{syn}} \cdot (V_a - V_{cm})\,.
\end{equation}
If $V_{mem}(t_{pre})<\theta_V$, then learn\_up is low and learn\_down is high, which results in the downward jump size of 
\begin{equation}
b = \frac{C_{hebb}}{C_{syn}} \cdot (V_b - V_{cm})\,.
\end{equation}

In order to reduce the number of control voltages, single-ended input voltages are provided. The resulting common mode offset, caused by this asymmetry, is compensated by the SC CMFB circuit.

The "stop learning" feature described in Sec. \ref{sec_fusi_model} is handled by setting learn\_up resp. learn\_down to low using combinational logic (not shown). Therefore, the state of the calcium variable can be calculated externally in an FPGA, where the postsynaptic spike train is filtered by a low pass filter. The low pass filter output is then compared against the stop learning thresholds $\theta_{up}^l$, $\theta_{up}^h$, $\theta_{down}^l$ and $\theta_{down}^h$ and the two resulting binary signals for enabling learning in the up and down direction, respectively, are transmitted to the driver circuit. As an additional feature for testing we implemented a "learn force" mode where learn\_up and learn\_down can be set explicitly, similar to keeping the neuron membrane permanently elevated or depressed.

The comp signal, which is provided in the "Comparison" phase states whether the synapse is depressed (LTD) or potentiated (LTP). This binary output is used to scale the PSC generated by the presynaptic adaptation circuit (see "Weight Scaling \& Charge Transmission" in Fig. \ref{fig_circuit}). Therefore each synapse has two 4-bit weights for LTP and LTD stored in a RAM (see Fig. \ref{fig_overview}), which is chosen accordingly to the synapse state and transmitted to the weight scaling circuit. The scaling of the PSC is done via binary weighted capacitors, transferring charge to the neuron circuit. Additionally each synapse is selectable excitatory or inhibitory, which inverts the PSC voltage. Thus, inhibitory stop-learning synapses are also possible.

\subsubsection{Comparator Circuit}
\label{sec_comparator}
A circuit schematic of the comparator shown in Fig. \ref{fig_ltpcircuit} is depicted in Fig. \ref{fig_compcircuit}a. 
It consists of a preamplifier (see Fig. \ref{fig_compcircuit}b), which is inspired by \cite{dessouky00} and a simple dynamic latch circuit \citep{song95} shown in Fig. \ref{fig_compcircuit}c. This architecture has been chosen, because the dynamic latch circuit can have a high random offset voltage of up to \SI{20}{\milli\volt}, caused by mismatch. The preamplifier raises the differential signal level to minimize decision errors, caused by this mismatch. The preamplifier is therefore equipped with an offset compensation (compare Sec. \ref{sec_offset_compensation}). At the output of the comparator circuit an SR-latch is connected, which stores the result until the next comparison.

\subsection{Measurement Setup and Characterization Methods}
\label{sec_char_time_constant}

\begin{figure}
\centering
\includegraphics[width=0.5\textwidth]{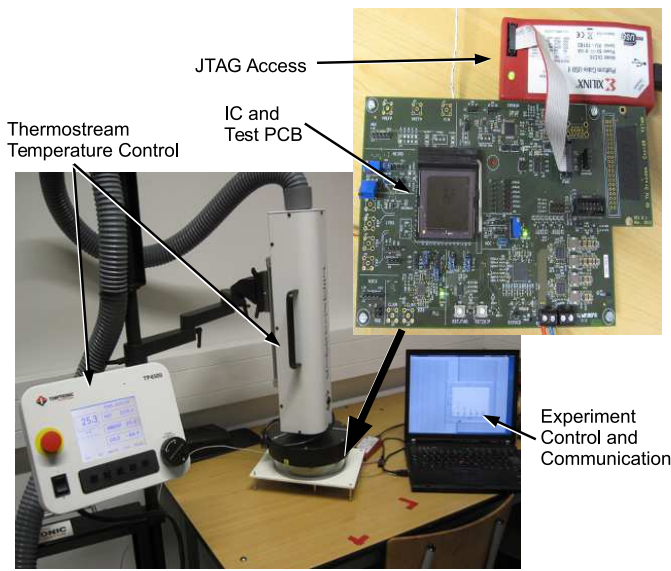}
\caption{\label{fig_thermo_setup} Setup for measurements with controlled temperature}
\end{figure}

As detailed in Sec. \ref{sec_design_system}, the entire system is ratiometric with respect to the clock frequency. That is, the system clock can be scaled so that the neuromorphic system operates anywhere from biological real time up to a factor 100 faster. As operation at biological real time is the most challenging in circuit terms as well as the most interesting in terms of computation, real-time operation was used for the measurements in this paper.
The corresponding clock frequency is \SI{3.3}{\mega\hertz}, generated by a configurable clock divider from the \SI{330}{\mega\hertz} central system clock.
At this frequency, the synaptic matrix update period is \SI{0.62}{\milli\second} (compare Sec. \ref{sec_design_system}).

The measurements of the presynaptic adaptation are carried out at the temperatures indicated by using the temperature controlled setup shown in Fig. \ref{fig_thermo_setup}. The IC package is held at the adjusted temperature with ca. \SI{\pm 2}{\degreeCelsius} deviation. 
The output of the presynaptic adaptation can be measured either via tracing the PSC time course from one of the analog test outputs or indirectly by monitoring the spike output of a connected neuron.
Directly measuring the PSC voltage via an oscilloscope is well-suited for detailed short-time measurements, which we used to verify correct operation of the circuitry.
For reducing noise in this case, the aquired waveform data was averaged over time bins of \SIrange[range-units=single,range-phrase = -- ]{0.1}{0.3}{\milli\second}.

Direct oscilloscope measurements are less practical for automatic extraction of a multitude of time constants.
For this case, we used the following purely spike-based protocol:
The adaptation state is probed by sending an input spike and counting the number of output spikes in reaction.
For getting a reasonably strong response, the synaptic weight and the PSC scaling voltage are set to their maximum values.
Setting the membrane time constant to a high value as well, the number of output spikes per input spike is approximately linearly dependent on the PSC amplitude.
For the measurements, we only activated depression, so that the PSC amplitude of a spike directly resembles the current state of the depression variable.
For each time constant measurement, the depression variable is charged by initially applying 10 spikes.
Afterwards, the adaptation strength is set to zero, so that the depression variable relaxes back to its resting state.
This relaxation is monitored by continuously probing the state with input spikes.
From the relaxation time course, the time constant is extracted by calculating the best-fitting (smallest root mean squared error) exponential function, with amplitude and time constant as free parameters.
Results are averaged over 10 repetitions.

The measurements of the stop learning synapses are carried out at ambient temperature, i.e. no special measures for chip cooling are taken.
\section{Results}

\subsection{Basic Operation of the Presynaptic Adaptation}
\label{sec_results_markram_basic}

\begin{figure}
\centering
\includegraphics[width=0.5\textwidth]{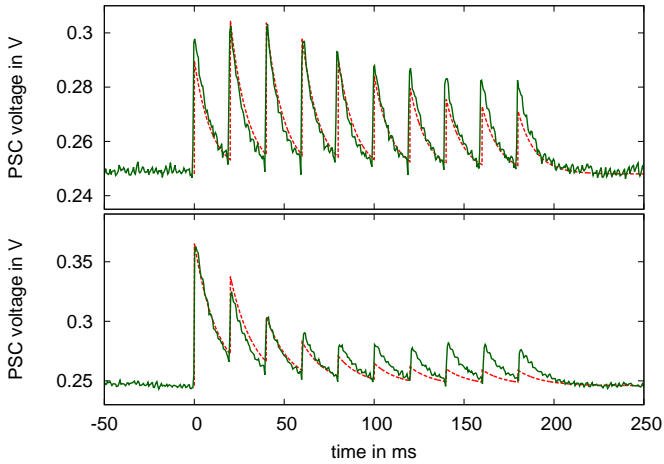}
\caption{\label{fig_presyn_depr_facil} PSC voltage traces of a simultaneously facilitating and depressing (top), and of a depressing (bottom) synapse when stimulated with 10 spikes at 50Hz rate. Configuration parameters: top: $\tau_u=300$ms, $\tau_R=300$ms, $\tau_\mathrm{PSC}=10$ms, $U=0.29$, $\alpha=0.5$, bottom: $\tau_u=10$ms, $\tau_R=490$ms, $\tau_\mathrm{PSC}=13$ms, $U=0.96$, $\alpha=0.5$. The nominal time courses for the PSC voltages with these parameters and fitted amplitudes are drawn as dashed lines.}
\end{figure}

For evaluating the presynaptic adaptation performance, we stimulated a presynaptic circuit with a regular spike train for two different adaptation types, as shown in Fig. \ref{fig_presyn_depr_facil}.
We chose a parameter set for combined facilitation and depression to demonstrate correct operation of the circuit as a whole, and a setting for a depressing synapse, where the depression variable dominates the behaviour.
The latter case is used for assessing the correct reproduction of long time constants in the next section.

Figure \ref{fig_presyn_depr_facil} also shows ideal time courses for the implemented model with the same parameters and fitted amplitude and offset.
The measurements agree well with these nominal curves even without calibrating any parameters.
They differ mainly in the adaptation strength, i.e. in the ratio between highest and lowest PSC amplitude, which is smaller in the measured curves.
This may be caused by time constants being too small, or by charge injection effects, resulting in voltage offsets during updates of the adaptation variables at incoming spikes.

\subsection{Characterization of the Presynaptic Adaptation Time Constants}
\label{sec_results_time_constants}

\begin{figure}
\centering
\includegraphics[width=0.5\textwidth]{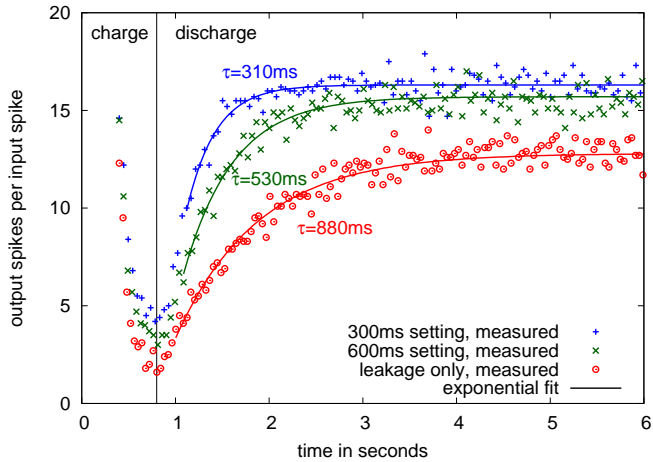}
\caption{\label{fig_rate_trace}Measured time courses of input-output gain for one presynaptic adaptation circuit at 40\degree C with 300 ms, 600 ms and leakage only settings. Time course until 0.8 s is the charging of the depression, following, the synapse relaxes back to its steady state with the depression time constant.}
\end{figure}

Fig. \ref{fig_rate_trace} shows traces over different time constant settings for one presynaptic adaptation circuit.
The time course of the depression relaxation for nominal settings as well as with only leakage present can be faithfully fitted by an exponential function, allowing for calculation of the depression time constant. 

\begin{figure}
\centering
\includegraphics[width=0.5\textwidth]{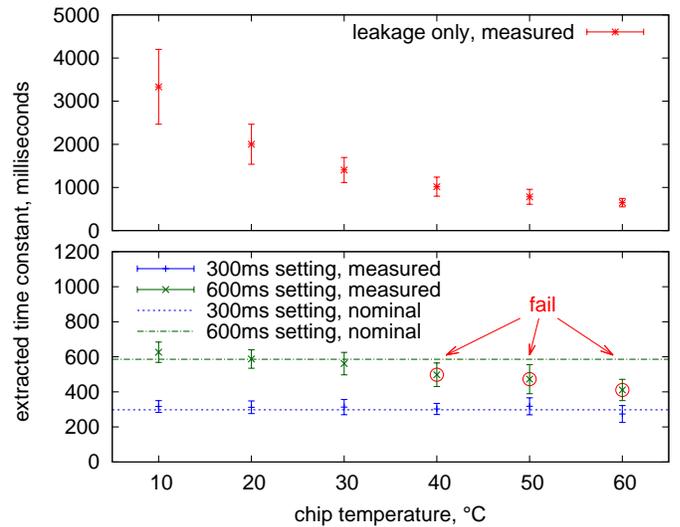}
\caption{\label{fig_tau_ensemble}Mean and standard deviation (error bars) of extracted time constants over 16 presynaptic adaptation circuits of four separate ICs. Shown is the measured time constant for a setting of infinity (upper part, i.e. the equivalent time constant if just leakage is active) and two configured time constants (nominal 600 and 300ms) for the presynaptic adaptation circuit of Fig. 3.}
\end{figure}

Measured time constants of 16 adaptation circuits from 4 chips are shown in Fig. \ref{fig_tau_ensemble}. The values are well-controlled in the configurable range up to \SI{300}{\milli\second} at all temperatures with sigma less than 15\% and the mean within 20\% of the nominal setting. The same is true for the 600~ms setting up to 30\degree C. Above that, the leakage influence causes the measured mean to be at least one sigma outside the nominal, which constitutes our fail criterion.

Using the infinite setting for the depression time constant, i.e. there are no decay switching events, this leakage can be measured, see upper plot in Fig. \ref{fig_tau_ensemble}.
As expected, it is highly temperature-dependent.
For temperatures of 30\degree C and below, all measurements are above 1 second, so that time constants up to this value are feasible at room temperature if the controlled leakage, i.e. the switching frequency of the decay process, is further decreased compared to the 600~ms setting.
As described in Sec. \ref{sec_leakage_reduction}, a time constant of 600~ms corresponds to a leakage resistance of 8~TOhm.
This value increases to a minimum of 13~TOhm for time constants of 1 second or above.
These high resistances demonstrate the effectiveness of the employed leakage reduction techniques.

The measurements show that time constants of several seconds are possible at temperatures below 30\degree C.
As the time constants caused by intrinsic leakage show a larger spread for these temperatures, individual calibration of the switching frequency for the leakage mechanism may be required to still achieve well-controlled time constant values.
Nevertheless, for the envisaged time constant range up to 600~ms of the design, the measurements demonstrate correct resemblence of time constant values at room temperature, so that all further measurements were performed without any special measures for temperature control.
\subsection{Characterization of the Bistable Stochastic Synapse}
\label{sec_results_synapse}

\begin{figure}
\centering
\includegraphics[width=0.5\textwidth]{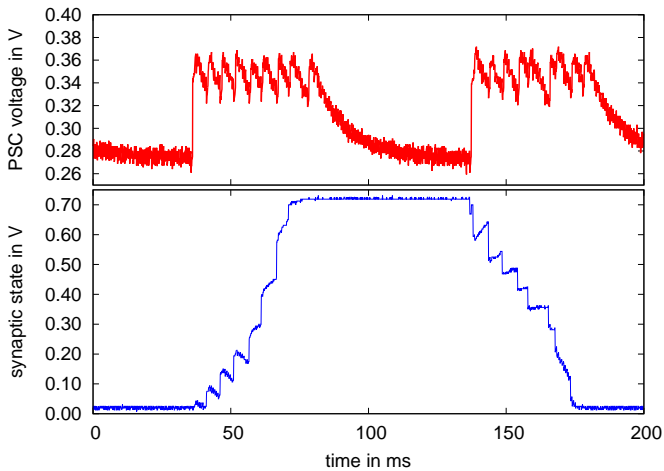}
\caption{\label{fig_fusi_synapse} (upper diagram) Measured PSC waveform of a \SI{200}{\hertz} presynaptic spike train with 12 pulses; (lower diagram) synapse state of stochastic stop learning synapse, with forced transition from depressed to potentiated state and back.}
\end{figure}

In this section, results for the SC implementation of the stop-learning synapse are given. As detailed in Sec. \ref{sec_fusi_circuit}, a force bit can be set that forces the synapse to transition from potentiated to depressed state or vice versa. That is, Eqn. \ref{eq_fusi_a} resp. Eqn. \ref{eq_fusi_b} are forced to always employ $a$ or $b$, similar to setting $V_{mem}(t)$ either to a constant high or low value. A presynaptic spike train of 12 spikes is then applied to the synapse, as shown in the upper diagram of Fig. \ref{fig_fusi_synapse}.  

From the lower diagram of Fig. \ref{fig_fusi_synapse}, it can be observed that the synapse reaches a stable potentiated state (at ca. 0.7~V) or a depressed state (at \SI{0}{\volt}). For the transition at \SI{50}{ms}, the force bit activates only $a$, forcing the synapse to become potentiated. Conversely, at \SI{150}{\milli\second}, only b is active, the synapse becomes depressed. Between presynaptic events, the curve shows that $\alpha$ and $\beta$ draw the synapse back to one of its stable states, according to the synapse state being above or below $\theta_X$ (set at half way between the two stable states, see also Eqn. \ref{eq_fusi_alpha} resp. \ref{eq_fusi_beta}).

\begin{figure}
\centering
\includegraphics[width=0.5\textwidth]{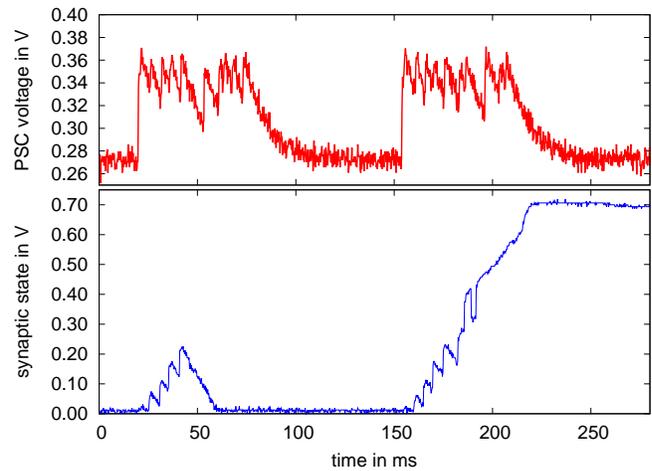}
\caption{\label{fig_fusi_synapse_sl} (upper diagram) Measured PSC waveform of presynaptic spike train, both packets 12 pulses, \SI{200}{\hertz}; (lower diagram) synapse state of stochastic stop learning synapse, with forced transition from depressed to potentiated state. The first transition is aborted due to activation of stop learning after 6 pulses, i.e. at a point where the synapse state is not above $\theta_X$ and thus gets drawn back to the depressed state. For the second transition, stop learning is activated after 8 pulses.}
\end{figure}

To test the stop learning functionality expressed in our implementation by the two stop learning bit flags (see Sec. \ref{sec_fusi_circuit}), a second experiment is carried out. The packet of 12 presynaptic spikes is split in two parts which are sent immediately after each other, see the corresponding PSC voltage in the upper diagram of Fig. \ref{fig_fusi_synapse_sl}. Starting from the depressed state, the force bit activates $a$, but after the first part of the presynaptic spike packet, which contains 6 pulses, the stop learning bit for $a$ is activated. This causes the last 6 pulses to be discarded in terms of synaptic state modification, i.e. only $\beta$ is active which draws the synapse back down to the depressed state. 

At \SI{150}{\milli\second}, this experiment is repeated, but the stop learning is activated after 8 pulses. This is sufficient to push the synapse above $\theta_X$, i.e. $\alpha$ becomes active which draws the synapse state to the potentiated state, even though the last 4 presynaptic pulses are again discarded because of the activated stop learning. Thus, overall functionality of the stochastic stop learning synapse is confirmed. In this experiment, the stop learning was set explicitely. As stated in Sec. \ref{sec_fusi_circuit}, the future backplane for a multi-chip system will compute the Calcium variable externally on an FPGA based on the output spike rates \citep{brader07}, setting the stop learning bits dynamically based on the Calcium state.

Please note that we are only showing the internal synaptic state transitions. For the overall network dynamics, the state change means a switch between the 4~bit potentiated and 4~bit depressed weights (compare Sec. \ref{sec_fusi_circuit}). Thus, while learning induction is in the form of the one bit decision of the original stop learning synapse \citep{brader07}, the expression of the synaptic learning can be individual for each synapse, adding significantly to network richness compared to the global settings for potentiated and depressed synapses in other implementations of this plasticity rule \citep{indiveri06}. This capability for individual weights could also be exploited for implementations of the Neural Engineering Framework \citep{eliasmith03} on our neuromorphic system. A 4~bit weight resolution plus the capability for setting each synapse excitatory or inhibitory should be sufficient for sophisticated population-based signal processing \citep{mayr14b}, compare also the results achieved for 58 neurons with 4~bit synaptic weights in \citep{corradi14}.

\subsection{Overall Results}
\label{sec_results_overall}

Table \ref{tab_characteristics} details the major characteristics of the neuromorphic system. Its power budget is competitive with recent power-optimized digital or analog neuromorphic systems of similar size \citep{indiveri06,seo12}. The digital part includes 0.45~mW static power draw which is mainly due to the other components on this test chip, so putting the neuromorphic system on a chip by itself would improve power consumption by about 23$\%$ at biological real time operation. The current clocking setup features a constant-frequency PLL \citep{hoeppner13} and a clock divider, which draw constant power irrespective of the speed up factor. To save power, this could be replaced with a variable-frequency PLL with frequency-dependent power draw \citep{eisenreich09}.

Plasticity models with time constants up to seconds have been shown for this SC implementation in 28~nm. Thus, reliable, controlled behaviour fully in keeping with biological real time operation is possible. The efficacy of our chosen method for low-leakage capacitive state holding has been proven, with detailed analysis of the effect of temperature on achievable time constants. The characterization of the presynaptic time constants employs the entire signal pathway of the system (compare Fig. \ref{fig_overview}), showing complete overall functionality. 

Table \ref{tab_comparison} gives a comparison with other current implementations of presynaptic adaptation and/or synaptic plasticity. The synapse area of our implementation is among the lowest, with only the static 1~bit synapse of a digital synaptic array smaller in size.  Especially, compared to fully analog implementations of stop learning \citep{indiveri06}, the SC approach and agressive scaling for the various capacitances allow an implementation of stop-learning that benefits from the technology shrink. As can be seen from the faithfulness of model replication in SC, this scaling can be achieved without compromising functional richness and accuracy. 
When accounting for technology node, the area consumption of the presynaptic adaptation is larger than e.g. \cite{bartolozzi07} or \cite{schemmel10}. This is due to the fact that our presynaptic adaptation aims at a very faithful reproduction of the model of \citep{markram98}, necessitating complex, multi-stage computational circuits (see Fig. \ref{fig_circuit}). Specifically, our implementation is the only one offering concurrently operating facilitation and depression.

The shown architecture always connects an input via synapses to all neurons, corresponding to an all-to-all connectivity.
This is the same architecture as used for example in memristive crossbar arrays \cite{alibart12,mayr12b}.
The main advantage of this architecture in our design is that it allows to implement all parts of the synapse circuit that depend on the input only once per synapse row.
This significantly reduces circuit area, reducing the synapse circuit to an analog storage element in our design.
The efficiency gain comes at the price of reduced flexibility concerning connection topologies.
All-to-all and comparable connection structures are well-suited, whereas sparse connectivity results in a high number of unused synapses in the matrix, making the architecture less efficient in this case, even when optimizing the mapping of networks to the hardware architecture \cite{galluppi12,mayr07a}.
To improve the efficiency, i.e. the fraction of utilized synapses, also for low connection densities, more presynaptic input circuits than synapse rows can be implemented, while synapses are made to choose between several inputs \citep{schemmel10,noack10}.
This would only slightly increase the complexity of the individual synapse circuits, while greatly increasing the flexibility of the architecture \citep{noack10}.

\begin{table}
\centering
\caption{\label{tab_characteristics}Characteristics of the presented SC neuromorphic system. All figures are for a speed-up of one, i.e. biological real time operation, if not stated otherwise.}
\footnotesize
\begin{tabular}{|p{2.5cm}|p{5cm}|}
\hline 
Technology & Global Foundries 28~nm SLP\\
\hline
Layout area for system & 460*430~$\mu\mathrm{m}^2$ neuromorphic comp., 600*600~$\mu\mathrm{m}^2$ overall (including DAC, RAM, etc.)\\
\hline
Clock frequency & 330~MHz (PLL), 3.3~MHz (neuromorphic components) \\
\hline
VDD analog & 1.0~V \\
\hline
VDD digital & 0.75~V \\
\hline
Power digital & 1.1~mW (speed-up 1) to 3.1~mW (speed-up 100) \\
\hline
Power analog (neuromorphic components) & 0.38~mW (speed-up 1) to 11.0~mW (speed-up 100) \\
\hline
Power analog (PLL) & 0.45~mW \\
\hline
Neuron model & LIAF \citep{rolls10} \\
\hline
Presynaptic adaptation & facilitation and depression \citep{noack12}\\
\hline
Synaptic plasticity &  stochastic synapse with stop learning \citep{brader07}\\
\hline
System characteristics &  128 presynaptic adaptation circuits, 8192 stochastic synapses, 64 LIAF neurons\\
\hline
\end{tabular}
\end{table}

\begin{table*}[ht]
\newcommand{\abcd}{p{1.7cm}}
\centering
\footnotesize
        \begin{tabular}{|p{1.5cm}|p{1.1cm}|p{1.1cm}|p{1.1cm}|p{1.0cm}|p{3.0cm}|p{1.0cm}|p{1.0cm}|p{3.0cm}|}
             \hline
              Ref. & Techn. & System area & Synapse area & Number of synapses & Synapse functionality & Pre-synapse area & Number of presynapses & Presynapse functionality\\
            \hline
             \cite{seo12,merolla12} & 45~nm & 4.2~mm$^2$ & 1.6~$\mu$m$^2$ & 262~k & 1-bit static synapses, set externally & -- & -- & not implemented \\
             \hline
             \cite{park14} & 90~nm & 16~mm$^2$ & 15~$\mu$m$^2$ & 262~k & log-domain conductance-based synapse, no plasticity & -- & -- & not implemented \\
	    \hline	
            \cite{mitra06,bartolozzi07} & 350~nm & 12~mm$^2$ & 1200~$\mu$m$^2$ & 8192 & stop learning & 1360~$\mu$m$^2$ & N.A. & short-term depression\\
            \hline
             \cite{schemmel10,schemmel14} & 180~nm & 50~mm$^2$ & 150~$\mu$m$^2$ & 115~k & STDP & 84~$\mu$m$^2$ & 14~k & Either short-term depression or facilitation\\
	\hline
           This work & 28~nm & 0.36~mm$^2$ &  13~$\mu$m$^2$ & 8192 & stop learning & 432~$\mu$m$^2$ &  128 & Concurrent short-term depression and facilitation\\

	\hline            
        \end{tabular}
    \caption{\label{tab_comparison}Comparison of the presented short- and long-term plasticity circuits with other implementations from literature.}
\end{table*}
\section{Discussion}
\subsection{Plasticity Models}

Results show faithful implementation of the chosen short-term plasticity model \citep{markram98}. The detailed reproduction of this model endows the neuromorphic system with a corresponding rich behavioral repertoire, which could be employed for e.g. reproduction of population dynamics in cultured neurons \citep{masquelier13} or simulation of short-term memory \citep{rolls13}. 

The long-term plasticity rule is also reproduced well, opening up a host of information-theoretic applications, such as studies of memory retention, information content or classification performance of a network \citep{brader07}. Other flavors of long-term plasticity rules could also be supported by our neuromorphic system. For instance, the faithful reproduction of neuronal waveforms evident in Fig. \ref{fig_presyn_depr_facil} and their excellent configurability in terms of the time window (Fig. \ref{fig_tau_ensemble}) could also be employed for a plasticity rule based on neuron and synapse waveforms such as \citep{mayr10b}, which aims at the replication of a wide range of biological plasticity experiments \citep{mayr10a}.

\subsection{Switched-Capacitor Neuromorphics}

Dating back to Carver Mead, subthreshold CMOS has been the mainstay of neuromorphic circuit design, as it offers the advantage of low power consumption, ion-channel like behaviour in CMOS devices and currents small enough to reach biological real time operation. 
However, such a fully analog implementation suffers from mismatch and leakage currents which are increasingly prevalent in deep submicron processes. In addition, the channel-to-transistor design philosophy means that this type of neuromorphic circuit consists largely of handcrafted circuits that depend crucially on the performance of each single transistor. Thus, porting a design between technology nodes essentially means a completely new design. 

Switched-capacitor neuromorphic circuits move from this device level philosophy to a building block approach, i.e. the required model behaviour is achieved with a combination of standard building blocks. SC is used as a mathematical framework to directly translate state-driven models to a mixed-signal realization. This keeps the neuronal states analog for biological veracity, while achieving significantly easier technology porting, as the circuit consists solely of standard building blocks such as amplifiers, switches and charge addition/subtraction. Representation of analog states at block level also eases implementation in deep submicron, as this takes advantage of the available device count for improved signal fidelity, while relying less on the characteristics of individual transistors. This building block approach allows agressive scaling of the active analog components, while the digital part of the SC circuits naturally scales with the technology node. Overall scaling is ultimately limited compared to a purely digital system by the largely invariant capacitor sizes, but is still significantly better than conventional, more device- and analog-centric neuromorphic approaches. As shown, this approach has enabled our SC system to deliver the same computational density as a purely digital neuromorphic system in a deep-submicron technology \citep{seo12}, while its power budget is on par with subthreshold circuits \citep{indiveri06}. When combined with deep submicron pixel cells \citep{henker07}, a sophisticated visual processing pyramid could be implemented \citep{serrano09,koenig02}. 

While SC makes neuromorphic circuits possible in principle in deep submicron, one major challenge is still the leakage currents. The leakage completely precludes subthreshold circuits, but it also affects the stored states of capacitors in SC technique, especially for the timescales necessary for biological real time operation. As shown, we have solved this general challenge for SC neuromorphic circuits with our low leakage switch architecture, reaching controllable time constants $>$100~ms at ambient temperature. 

\subsection{Nanoscale CMOS and Novel Devices}

Novel nanoscale devices, such as memristors, offer the possibility of very high density neuromorphic synaptic matrices  \citep{alibart12,ou13}. However, they need corresponding high-density neuronal driver circuits in CMOS. Moving neuromorphic circuits to deep-submicron technologies as outlined in this paper would provide this capability, i.e. very low footprint neuron driver and receiver circuits that generate analog waveforms for memristor synaptic matrices \citep{mayr12b}.

\section*{Disclosure/Conflict-of-Interest Statement}

The authors declare that the research was conducted in the absence of any commercial or financial relationships that could be construed as a potential conflict of interest.

\section*{Acknowledgement}
This work is partly supported by 'Cool Silicon', the 'Center for
Advancing Electronics Dresden' and the European Union 7th framework program, project 'CORONET' (grant no. 269459).

\bibliographystyle{spbasic}
\bibliography{titan_timeconstants}

\end{document}